\documentclass{ws-ijmpd}  
    
\date{\today}  
\def \be{\begin{equation}}
\def \bea{\begin{eqnarray}}
\def \eea{\end{eqnarray}}
\def \ee{\end{equation}}

\def \A {{\cal A}}
\def \st{\tilde{s}}
\def \h {\frac{1}{2}}
\def \W {{\cal W}}
\def \h {{\frac{1}{2}}}

\def \DW {\Delta_{\W}}
\def\lsim{\mathrel{\rlap{\lower4pt\hbox{\hskip1pt$\sim$}}
    \raise1pt\hbox{$<$}}}                
\def\gsim{\mathrel{\rlap{\lower4pt\hbox{\hskip1pt$\sim$}}
    \raise1pt\hbox{$>$}}}                

\begin{document}  
\markboth{S. Dhurandhar, H. Mukhopadhyay, H.Tagoshi and N. Kanda}
{Coherent vs coincidence search}
%
\catchline{}{}{}{}{}
%
\title{Coherent versus coincidence detection of gravitational wave signals from compact inspiraling binaries}
\author{S. Dhurandhar$^1$, H. Mukhopadhyay$^1$, H. Tagoshi$^2$ and N. Kanda$^3$}  
\address{ $^1$IUCAA, Postbag 4, Ganeshkind, Pune - 411 007, India.  \\ 
 $^2$Department of Earth and Space Science, 
Graduate School of Science, Osaka University, Osaka 560-0043, Japan.  
\\ $^3$Department of Physics, Graduate School of Science, Osaka City University, 
Osaka 558-8585, Japan. 
}  
  
\maketitle  

\begin{abstract}  

We compare two multi-detector detection strategies, namely, the coincidence and the coherent, for the detection of spinless inspiraling compact binary gravitational wave (GW) signals. The coincident strategy treats the detectors as if they are isolated - compares individual detector statistics with their respective  thresholds while the coherent strategy combines the detector network data {\it phase coherently} to obtain a single detection statistic which is then compared with a single threshold. In the case of geographically separated detectors, we also consider an {\it enhanced} coincidence strategy because the usual (naive) coincidence strategy yields poor results for misaligned detectors. For simplicity, we consider detector pairs having the same power spectral density of noise, as that of initial LIGO and also assume the noise to be stationary and Gaussian. We compare the performances of the  methods by plotting the \emph{receiver operating characteristic} (ROC) for the two  strategies. A single astrophysical source as well as a distribution of sources is considered. We find that the coherent strategy performs better than the two coincident strategies under the assumptions of stationary Gaussian detector noise. 

\end{abstract} 

\section{Introduction \label{SC:1}}  

Inspiraling binaries are one of the most promising sources for the first detection of gravitational waves (GW). The post Newtonian approximation methods accurately describe the phasing of the waveform - about a cycle in a wave train $\sim 10^4$ cycles long. This makes it amenable for matched filtering analysis. The best available estimates suggest that the expected number of neutron star (NS)-NS binary coalescence seen per year by ground based interferometers is $7.1 \times 10^{-3} - 0.12$ for initial detectors and $38-6.6\times 10^2$ for advanced detectors \cite{kalogera}. In recent years, a number of ground based detectors are producing sufficiently interesting sensitive data and analysis of network data is highly advisable. The advantages of multi-detector search for the binary inspiral is that, not only does it improve the confidence of detection, it also provides directional and polarisational information about the GW source. 

Two strategies currently exist in searching for inspiraling binary sources with a network of detectors: the coherent and the coincident. The coherent strategy involves combining data from different detectors phase coherently, appropriately correcting for  time-delays and polarization phases and obtaining a single statistic for the full network, that is optimized in the maximum likelihood sense. On the other hand, the coincident strategy matches the candidate event lists of individual detectors for consistency of the estimated parameters of the GW signal. However, the phase information is ignored and as also the detectors are considered in isolation.

The question arises as to which strategy performs better. On simple waveforms the analysis has been performed by Finn and Arnaud et al\cite{Finn,Arnaud}. Both these works have shown that the coherent strategy performs better than the coincident strategy. We consider here the astrophysically important source, namely, the inspiraling binary and report mainly the results of our work which has been described in detail in the papers \cite{paper1,paper2,paper3}. We compare the strategies by plotting the \emph{Receiver Operating Characteristic} (ROC) curves,  which is the plot of detection efficiency versus the false alarm rate. We broadly consider the two cases of co-located aligned detectors and geographically separated misaligned detectors. In the co-located case we further consider two subcases of (i) uncorrelated noise, and (ii) correlated noise. 

\section{The correlation statistic}

  For the inspiraling binary, in the Fourier domain we assume the spinless restricted post-Newtonian (PN) waveform ${\tilde h}^I (f)$ at detector $I$:
\be
\tilde{h}^I (f) = {\mathcal{N}} E^I f^{-7/6} \exp ~ i [\Psi (f; t_c, \delta_c, \tau_0, \tau_3) + 2 \pi f \Delta t^I] \,,
\label{eq:stilde}			   
\ee
where we take the phase $\Psi$ given by the 3 PN formula. The extended beam pattern functions \cite{PDB} $E^I$ encode the orientation and direction parameters of the source and detectors; the parameters $t_c, \delta_c$, are respectively, the time of coalescence, phase of coalescence; the quantities $\tau_0, \tau_3$ are the chirp time parameters which are independent functions of the two masses of the stars comprising the binary, the $\Delta t^I$ denote time-delays at the detector $I$ with respect to a fiducial detector and ${\cal N}$ is the amplitude depending on the masses and the distance to the source.  

In matched filtering, it is natural to define the scalar product $(a,b)$ of two real functions by,
\be
\left ( a, b \right ) = 2 \int_{f_l^I}^{f_u^I} df \ \frac{  \tilde{a}(f) \tilde{b}^*(f) \ + \ 
\tilde{a}^*(f) \tilde{b}(f)}{S_h (f)},
\label{eq:scalar}
\ee
where, we use the Hermitian property of Fourier transforms of real functions. $S_h (f)$ is the one sided power spectral density (PSD) of the noise which is assumed to be the same for all the detectors. Then the normalised templates $s_0, s_{\pi/2}$ corresponding to the intrinsic parameters $\vec \mu \equiv \{\tau_0, \tau_3 \}$ are defined via the equation, 
\be
\tilde{h}^I(f;\vec \mu_i, t_c, \delta_c) = \A^I (\st_0 (f; \vec \mu_i, t_c) \cos \delta_c  +  \st_{\pi/2} (f; \vec \mu_i, t_c) \sin \delta_c) \, .
\label{eq:quadrature}
\ee
We have the relation $\st_{\pi/2} (f; \vec \mu_i, t_c) = i \st_{0} (f; \vec \mu_i, t_c)$ where $\mu_i$ represents a grid point in the intrinsic parameter space and the normalisation of templates requires that the scalar products $(s_0,s_0) = (s_{\pi/2}, s_{\pi/2}) = 1$. We then define the complex correlation $C^I = c_0^I + i c_{\pi/2}^I$ where, the real correlations $c_0^I$ and $c_{\pi/2}^I$ are obtained by taking the scalar products of the data $x^I$ with $s_0$ and $s_{\pi/2}$ respectively; that is, $c^I_{0, \pi/2} = (s_{0, \pi/2}, x^I)$. Then the single detector statistic  for detector $I$ is just $\Lambda_I = |C^I|^2$. To decide detection in a given single detector $I$, $\Lambda_I$ is maximised over the template parameters $\vec \mu_i, t_c$ and compared with a preassigned threshold.

\section{Coincidence versus coherent detection}

For coincidence detection, the procedure is as follows:
\begin{itemize}
\item Choose the same threshold $\Lambda^*$ for the two detectors. 

\item Prepare two candidate event lists  such that $\Lambda_I  >  \Lambda^*, ~ I = 1, 2$. Look for pairs of candidate events, each candidate event coming from a different list, such that the sets of estimated parameters $t_c, \tau_0, \tau_3$ match - $|\Delta t_c | \leq \DW t_c, ~~ |\Delta \tau_0 | \leq \DW \tau_0, ~~ |\Delta \tau_3 | \leq \DW \tau_3$ where the $\Delta \lambda$ denotes the difference in the measured parameter $\lambda$, where $\lambda$ stands for any of the parameters $t_c, \tau_0, \tau_3$. The allowed error box is denoted by $\DW \lambda$. A quadratic sum of the noises is taken to determine the error box. We fix this box by performing simulations, so that the final probability of not losing an event is $0.97$ - on each parameter, we atmost allow 1 \% loss in events. For geographically separated detectors, for fixing the window size in $t_c$, the light travel time between the detectors is taken into account and added in quadratures to the errors due to noise in each detector.  
\end{itemize}

On the other hand, coherent detection involves combining data streams in a phase coherent manner so as to effectively construct a single, more sensitive detector. For the case of two misaligned detectors the network statistic is given by \cite{PDB}: 
\be
\Lambda = ||C||^2 = |C^1|^2 + |C^2|^2 =  (c^1_0)^2 + (c^1_{\pi/2})^2 + (c^2_0)^2 + (c^2_{\pi/2})^2 \,,
\label{twoindep}
\ee
where $C^I$ is the complex correlation of the $I$-th detector ($I$=1,2). For two aligned colocated detectors the statistic is different \cite{PDB} and is given by: 
\be
\Lambda = \h |C^1 + C^2|^2 \,.
\ee
For aligned colocated detectors we consider two subcases: (i) uncorrelated noise, (ii) correlated noise. 
  
\section{Results}
\subsection{Colocated aligned detectors}

\begin{figure}[hbt]
\begin{tabular}{cc}
\begin{minipage}{0.5\hsize}
\centerline{\psfig{file=roc_new.eps,width=5cm}}
\end{minipage}
\begin{minipage}{0.5\hsize}
\centerline{\psfig{file=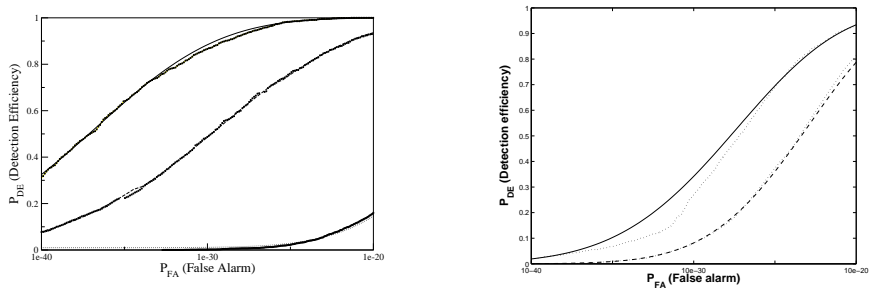,width=5cm}}
\end{minipage}
\end{tabular}
\caption{The ROC curves for single detector, coincident and coherent analysis for an injected signal of SNR=10. The solid line, dashed line and the dotted line correspond to the theoretical ROC curves for coherent, coincident and single detectors respectively. The left hand side figure corresponds to uncorrelated noise while the right hand one to correlated noise with $\epsilon_0 = 0.3$.}
\label{fig1} 
\end{figure}

 We first consider the case of co-located aligned detectors. This case would be of significance to the two LIGO detectors at Hanford and other similar topologies envisaged elsewhere in the future such as LCGT. We compare the performances of the two strategies by plotting the ROC curves for uncorrelated noise (left) and correlated noise (right) in Fig. (\ref{fig1}). For plotting these curves the false alarm and the detection probabilities must be computed as a function of the threshold $\Lambda^*$ and then plotted versus each other parametrically by varying the parameter $\Lambda^*$. The correlation parameter $\epsilon_0$ is taken as the weighted average of a frequency dependent correlation $\epsilon (f)$ where $\langle n_1 (f) n^{*}_2(f') \rangle = \h \epsilon (f)  S_h (f) \delta (f - f')$, $n_I (t)$ being the noise in the $I^{\rm th}$ detector. Details may be found in \cite{paper1,paper2}. While performing the simulations there are many subtleties such as estimating the number of independent templates, the error window size computation etc. the discussion of which we have omitted here although it is nonetheless important.  

It is clear from the figures that the coherent strategy is far superior in this case.

\subsection{Geographically separated misaligned detectors}

We now consider the case of geographically separated detectors which are then also normally misaligned. When the detectors are misaligned, the sky coverage for the usual coincidence detection is very poor which leads to intolerable false dismissal. Thus another coincidence strategy is devised which we call \emph{enhanced} coincidence. The usual coincidence strategy we then call \emph{naive} coincidence. Enhanced coincidence strategy is formulated as follows: 

\begin{itemize}
\item Choose a low threshold $\Lambda_0^*$ (we choose $\Lambda_0^* \sim 16$), and  prepare two candidate event lists  such that $\Lambda_I  >  \Lambda_0^*, ~ I = 1, 2$. 

\item Look for a pair of candidate events, the events coming from separate lists, such that the sets of estimated parameters match within the error-window. The procedure is the same as the co-located case, except for the parameter $t_c$ in which the distance between the detectors enters.

\item  Choose the final (high) threshold $\Lambda^* > 2 \Lambda_0^*$ and construct the final statistic $\Lambda = \Lambda_1 + \Lambda_2$ and register detection if $\Lambda > \Lambda^*$. 
\end{itemize}

\begin{figure}[hbt]
\begin{tabular}{cc}
\begin{minipage}{0.5\hsize}
\centerline{\psfig{file=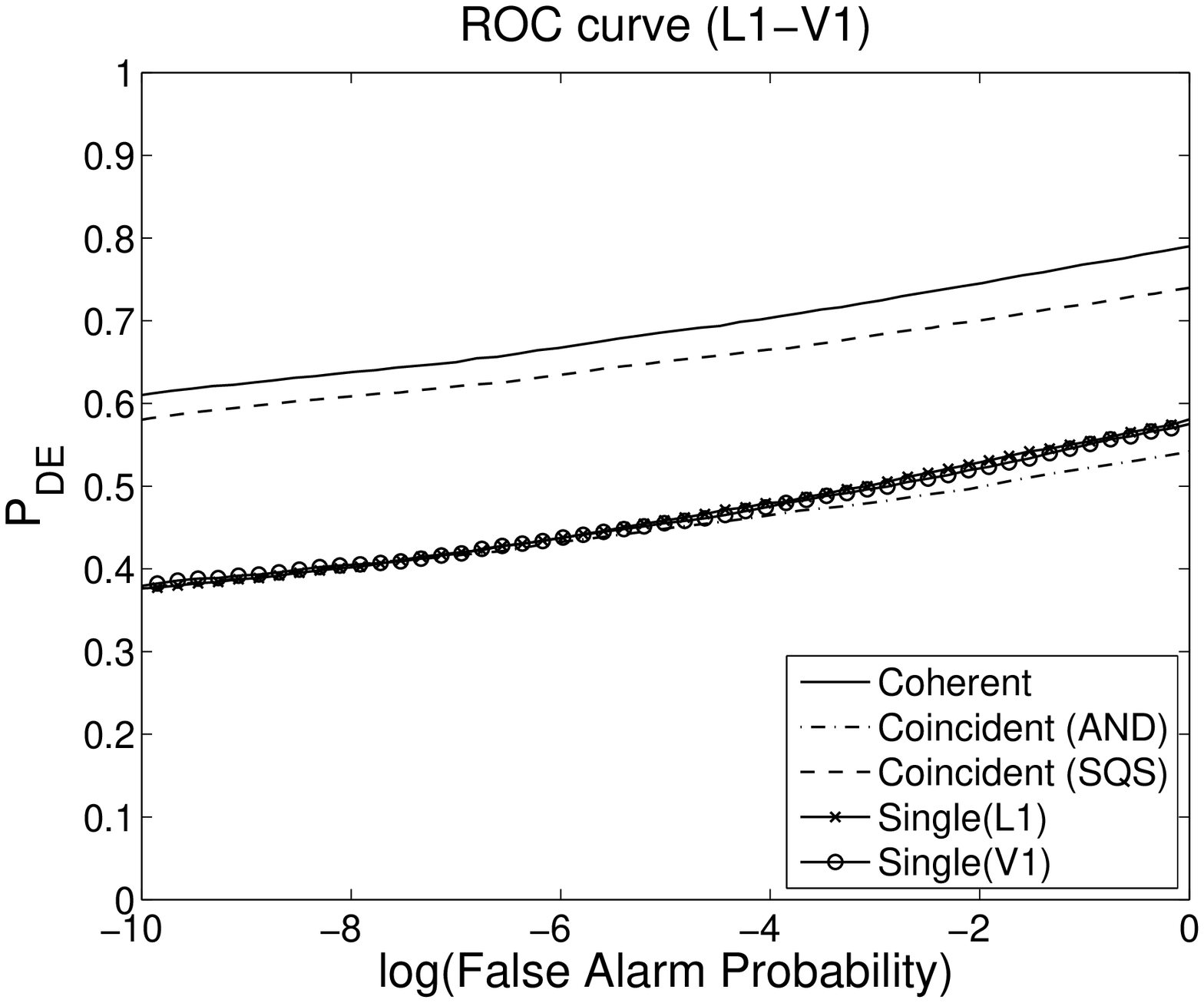,width=5cm}}
\end{minipage}
\begin{minipage}{0.5\hsize}
\centerline{\psfig{file=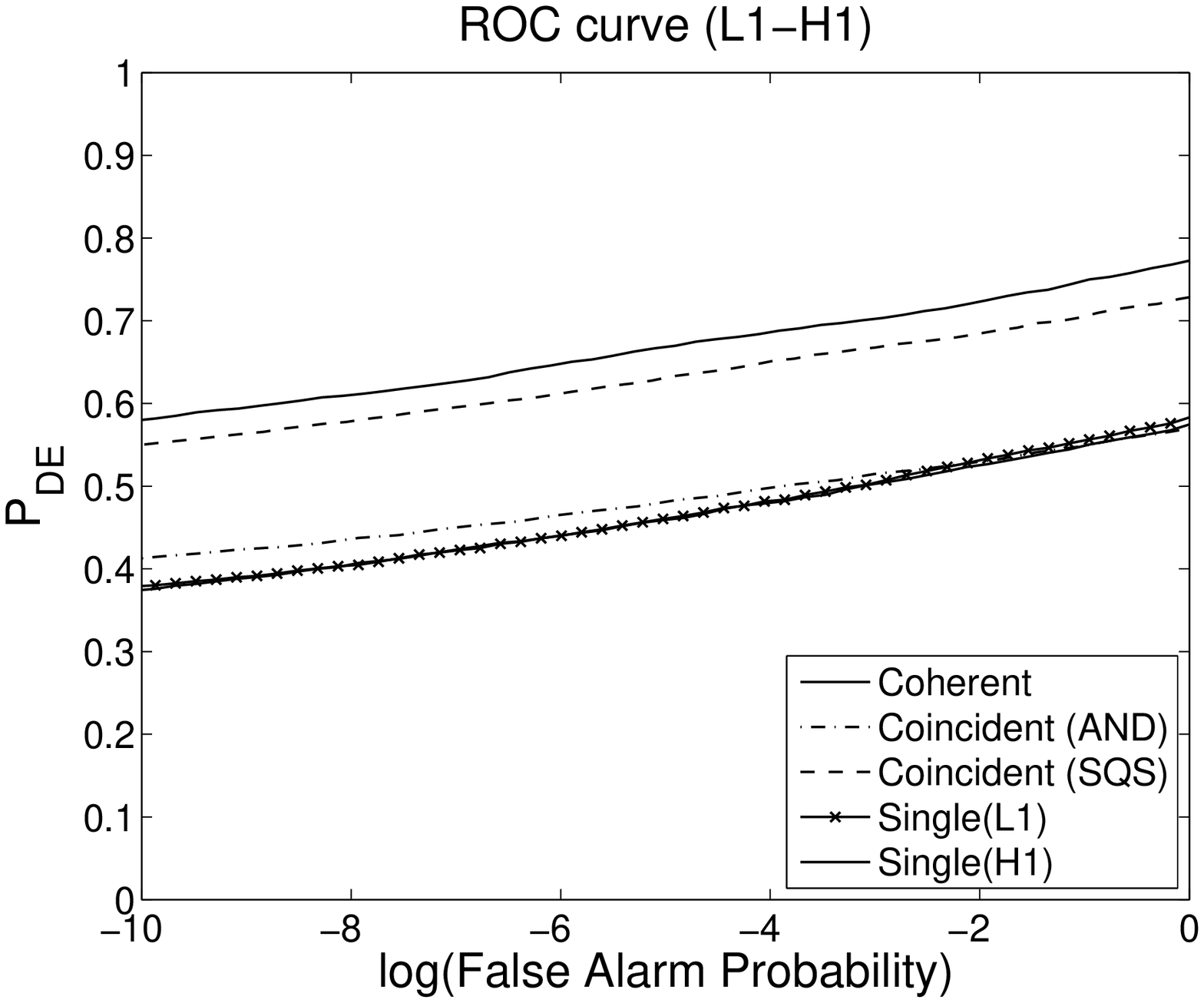,width=5cm}}
\end{minipage}
\end{tabular}
\caption{The ROC curves have been plotted for coherent, naive coincidence and enhanced coincidence for uniformly distributed sources. The left hand figure is drawn for the LIGO, Livingston and Virgo pair of detectors, while the right hand figure corresponds to the LIGO Livingston and LIGO Hanford detector pair.} 
\label{fig2}
\end{figure}

Note that although this statistic looks formally like the coherent case, the mass parameters for the templates in the two detectors do not have to be the same (they however must be close enough so that they lie in a error window). Thus this is not matched filtering while coherent detection is. In this strategy the sky coverage is better than naive coincidence. As seen from the Fig. (\ref{fig2}), coincidence strategy performs far better than the naive coincidence strategy, but we see that the detection probability for the coherent strategy is still superior by around 5\% for the same false alarm rate. 

\section{Concluding remarks}

Although the coherent strategy is superior to coincident strategies,
the difference between the coherent and enhanced coincident strategies is small.
Only a relative improvement of about 5\% in the detection probability is obtained with the coherent strategy. One may ask whether there is any practical advantage in using the coherent strategy in the case of two misaligned detectors. Note however that the coherent method is not so computationally expensive compared with two coincident methods, since we do not take cross correlation of two detectors' data in the coherent strategy. Thus overall, we conclude that the coherent strategy is a good detection method.
\par
However, the above results assume stationary Gaussian noise. But we know that the current real data are neither stationary nor Gaussian. In coincidence detection, the requirement of the consistency of estimated parameters in an error window acts as a powerful veto \cite{BSS_BFS} to veto out fake events generated from non-Gaussian noise. On the ther hand in coherent detection as yet no such obvious veto has been developed. This however does not rule out the possibility that a powerful veto cannot be constructed for the coherent strategy. In future we propose to work on this aspect of the problem. Perhaps a judicious combination of the two methods might be an effective way of dealing with this problem.  

\section*{Acknowledgments}

The authors would like to thank the DST, India and JSPS, Japan for the Indo-Japanese cooperative programme for scientists and engineers under which this work has been carried out.

\end{document}